Correspondence on "Electronic correlations in the iron pnictides"

I.I. Mazin, Naval Research Laboratory

In a recently published Letter[1] the authors have measured the Drude spectral weight, which is the same as the squared plasma frequency or as ($n/m$), where $n$ is the effective carrier density and $m$ is the effective mass, in LaFePO. They have convincingly shown that the experimentally measured spectral weight is nearly twice smaller than the one calculated in the standard density-functional based theory. They compared this to the mass renormalization obtained in a similar way for strongly correlated cuprates (by a factor of three or larger), and to the mass renormalization in itinerant metals near a ferromagnetic quantum critical point, such as Sr ruthenates. Quite insightfully, they point out that mass renormalization in these two examples has rather difficult physical nature, and that the iron pnictides are similar to itinerant ruthenates and dissimilar to near-Mott-insulators, cuprates. So far these conclusions are well supported by the experiment.

The authors then go one step further and conclude that the discrepancy between the experiment and the band structure calculations must be *entirely* ascribed to correlation effects (meaning, of course, an effect of spin fluctuations, and not the local Mott-Hubbard correlations). This would have been perfectly logical in a one-band system like cuprates, but not in a semimetal, and pnictides are classical semimetals. In semimetals the calculated plasma frequency strongly depends on the overlap of the hole of the electron bands (see Fig. 1). Incorrect relative positions of the hole and electron bands (such errors are very common even in simple materials, and there are indications that shifts of the order of ±100 meV may be necessary in LaFePO[2]) can change the calculated plasma frequency a lot. For example, density functional calculations predict germanium to have a finite bands overlap, *i.e.*, to be metallic – an infinite renormalization, if interpreted in the same vein as the discrepancy discovered by Qazilbas *et al.*!

In Fig.2 I show the calculated Drude spectral weight for LaFePO in the experimental crystal structure as a function of energy, separately for the hole and electron bands. Obviously, shifting the electron bands up and the hole bands down greatly reduces the total spectral weight. The inset in this Fig. shows the dependence of the spectral weight at the Fermi energy as a function of the shift. One can see that shifting by ±100 meV results in a 40% reduction of the spectral weight, leaving only 20% to a spin-fluctuations induced renormalization.

I want to emphasize that this fact does not invalidate the conclusions of Ref. 1; they may be correct, or may be not. It just appears that from these data, that is to say, from the observed discrepancy between the optical measurements and the calculations, one cannot draw conclusions about the strength of many body effects. They may be the leading source of the renormalization (in which case renormalization of the spectral weight is equivalent to the effective mass renormalization in the $n/m$ ratio), or the discrepancy may be partially or largely due to misalignment of the hole and electron bands in the calculations (which amounts to a renormalization of $n$).

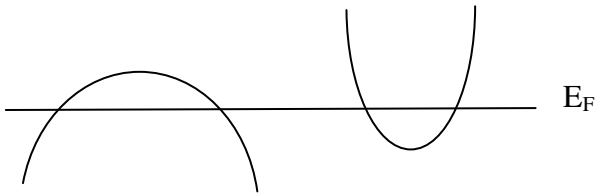

Figure 1: Schematic representation of a semimetallic band structure.

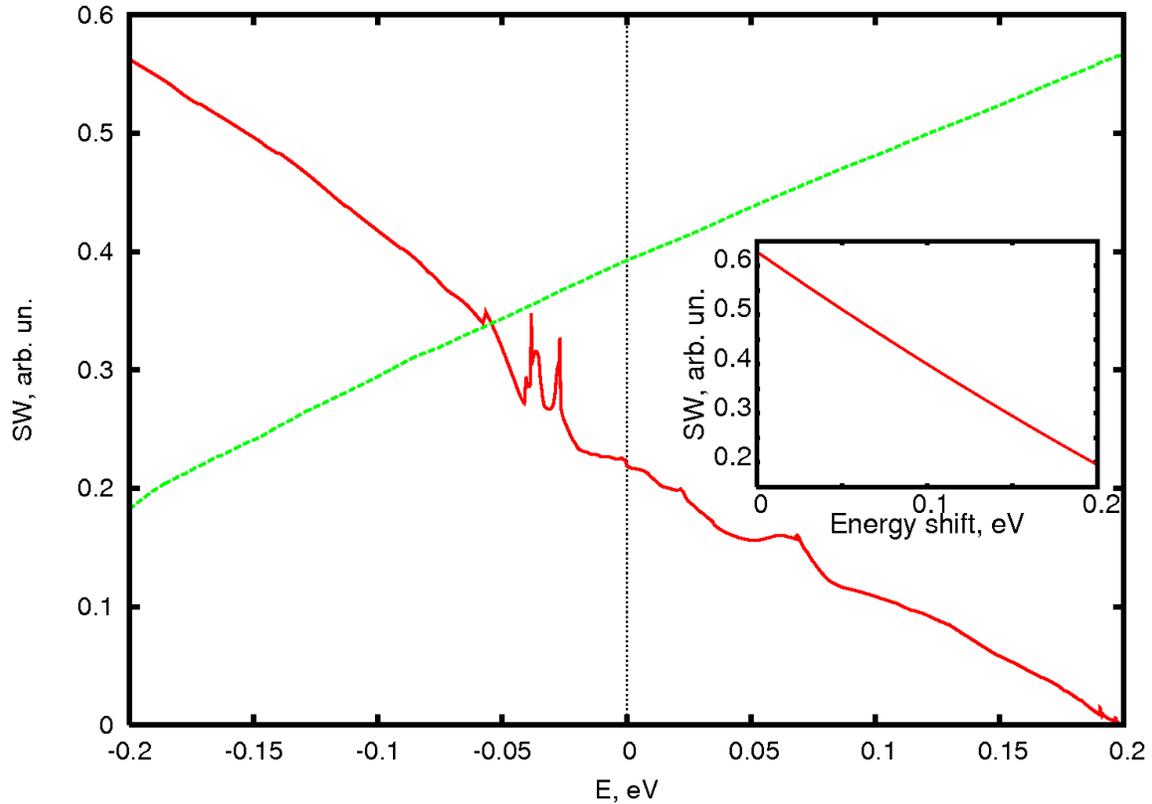

Figure 2. Calculated Drude spectral weight for LaFePO as a function of the Fermi level position, for hole and electron bands separately. Inset: Reduction of the spectral weight if the electron and the hole bands are given an energy shift of $\pm\Delta E$.

[1] M. M. Qazilbash, J. J. Hamlin, R. E. Baumbach, Lijun Zhang, D. J. Singh, M. B. Maple and D. N. Basov, *Electronic correlations in the iron pnictides*, Nature Phys. **5**, 647-650 (2009).
[2] A.I. Coldea, J.D. Fletcher, A. Carrington, J.G. Analytis, A.F. Bangura, J.-H. Chu, A. S. Erickson, I.R. Fisher, N.E. Hussey, and R.D. McDonald, *Fermi surface of a ferrooxypnictide superconductor determined by quantum oscillations*. Phys. Rev. Lett. **101**, 216402 (2008).